\documentstyle[12pt,psfig]{article}

\makeatletter
\def\ps@myheadings{\let\@mkboth\@gobbletwo
 \def\@oddhead{\hfil\rightmark}%
 \def\@oddfoot{\hfil\rm\thepage\hfil}%
 \def\@evenhead{\@oddhead}%
 \def\@evenfoot{\@oddfoot}\def\sectionmark##1{}\def\subsectionmark##1{}}
\makeatother

\newcommand{\nc}{\newcommand}

\nc{\be}{\begin{equation}}
\nc{\ee}{\end{equation}}
\nc{\beq}{\begin{equation}}
\nc{\eeq}{\end{equation}}
\nc{\beqa}{\begin{eqnarray}}
\nc{\eeqa}{\end{eqnarray}}
\def\nn{\nonumber}
\def\al{&&\!\!\!\!\!\!\!\!\!}
\def\a{&\!\!\!}

\author{\large  Faheem Hussain$^1$, Roberto Iengo$^2$, Carmen N\'u\~nez$^3$ 
and \\ Claudio A. Scrucca$^{2,4}$\\ \\
{\normalsize \em $^1$ International Centre for Theoretical Physics, Trieste, 
Italy}\\
{\normalsize \em $^2$ International School for Advanced Studies and INFN, Trieste, 
Italy}\\
{\normalsize \em $^3$ Instituto de Astronom\'{\i}a y F\'{\i}sica del Espacio 
(CONICET), Buenos Aires, Argentina}\\
{\normalsize \em $^{4}$ Institut de Physique Th\'eorique, Universit\'e de 
Neuch\^atel, Switzerland}}

\title{
\vskip -30pt
\normalsize
\begin{flushright}
SISSA REF 141/97/EP
\end{flushright}
\vskip 20pt
\Huge Interaction of D-branes on orbifolds and massless 
particle emission}

\date{}

\begin{document}

\maketitle

\thispagestyle{myheadings}

\begin{abstract}

We discuss various D-brane configurations in 4-dimensional orbifold 
compactifications of type II superstring theory which are point-like 
0-branes from the 4-dimensional space-time point of view. We analyze their
interactions and compute the amplitude for the emission of a massless NSNS
boson from them, in the case where the branes have a non vanishing relative
velocity. In the large distance limit, we compare our computation to the 
expected field theory results, finding complete agreement.

\end{abstract}

\begin{center}

Talk presented by Claudio A. Scrucca

\end{center}

\section{Introduction and summary}

We discuss various D-brane configurations in generic orbifold 
compactifications which are 0-branes from the 4-dimensional 
space-time point of view, but can have extension in the compact directions. 
More precisely, two cases turn out to be particularly interesting; the 0-brane
of type IIA and the 3-brane of type IIB.

The dynamics of these D-branes is determined by a one loop amplitude 
which can be conveniently evaluated in the boundary state formalism 
\cite{Polcai,Call}. 
In particular, one can compute the force between two D-branes moving with 
constant velocity, extending Bachas' result \cite{Bachas} to 
compactifications breaking some supersymmetry \cite{Hins1}.

Analyzing the large distance behavior of the interaction and its velocity 
dependence, it is possible to read the charges with respect to the massless 
fields, and relate the various D-brane configurations to known solutions 
of the 4-dimensional low energy effective supergravity.

Finally, we will discuss the emission of massless NSNS states
from two interacting D-branes \cite{Hins2}. 
The correlators that are involved have twisted boundary conditions because 
of the non zero velocity of the branes, but they can be systematically 
computed in a natural way using again the boundary state formalism. 
We will then briefly outline the large distance behavior of the string 
amplitude and its field theory interpretation.

\section{Interactions on orbifolds}

Consider two D-branes moving with velocities $V_1 = \tanh v_1$, 
$V_2 = \tanh v_2$ (say along 1) and transverse positions $\vec Y_1$, 
$\vec Y_2$ (along 2,3). 

\begin{figure}[h]
\centerline{\psfig{figure=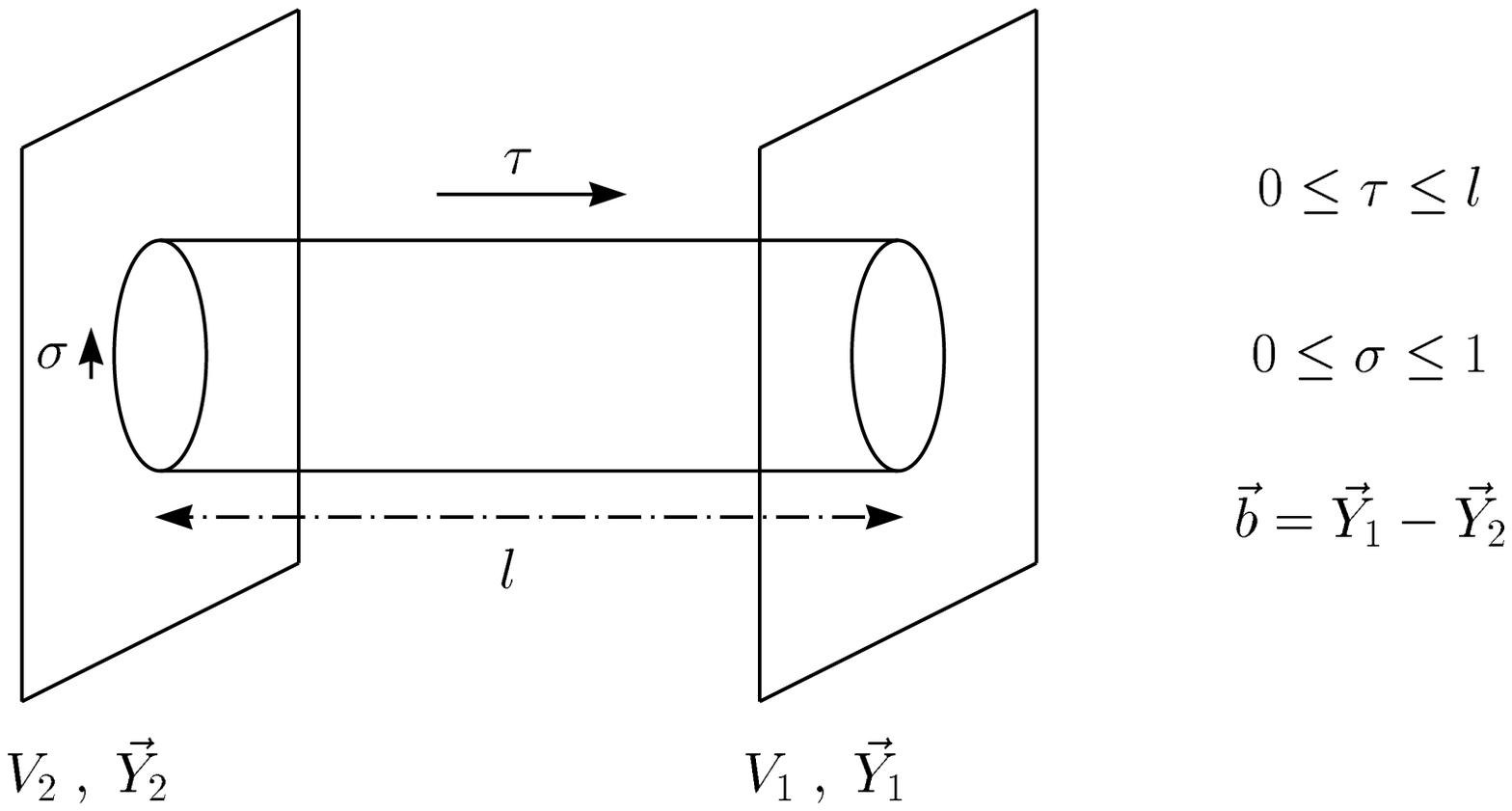,width=300pt}}
\label{fig1}
\end{figure}

The potential between these two D-branes is given by the cylinder 
vacuum amplitude and can be thought either as the Casimir energy 
stemming from open string vacuum fluctuations or as the interaction 
energy related to the exchange closed strings between the two 
branes. The amplitude in the closed string channel
$$
{\cal A}=\int_{0}^{\infty}dl \sum_s <B,V_1,\vec Y_1|e^{-lH}|B,V_2,\vec Y_2>_s
$$
is just a tree level propagation between the two boundary states, which are 
defined to implement the boundary conditions defining the branes. 

There are two sectors, RR and NSNS, corresponding to periodicity and 
antiperiodicity of the fermionic fields around the cylinder, and after
the GSO projection there are four spin structures, R$\pm$ and NS$\pm$,
corresponding to all the possible periodicities of the fermions on the 
covering torus.

In the static case, one has Neumann b.c. in time and Dirichlet b.c. in
space. The velocity twists the 0-1 directions and gives them rotated b.c. 
The moving boundary state is most simply obtained by boosting the static one 
with a negative rapidity $v=v_1-v_2$ \cite{Billo}.
$$
|B,V,\vec Y> = e^{-ivJ^{01}}|B,\vec Y> \;.
$$
In the large distance limit $b \rightarrow \infty$ only world-sheets with
$l \rightarrow \infty$ will contribute, and momentum or winding in the 
compact directions can be safely neglected since they correspond to 
massive subleading components.

The moving boundary states
\beqa
\al |B,V_1,\vec Y_1>=\int\frac{d^{3}\vec k}{(2\pi)^{3}}
e^{i \vec k \cdot \vec Y_1} |B,V_1> \otimes|k_B> \;, \\
\al |B,V_2,\vec Y_2>=\int\frac{d^{3}\vec q}{(2\pi)^{3}}
e^{i \vec q \cdot \vec Y_2}|B,V_2> \otimes|q_B> \;,
\eeqa
can carry only space-time momentum in the boosted combinations 
\beqa
\al k_B^\mu = (V_1 \gamma_1 k^1, \gamma_1 k^1, \vec k_T) = 
(\sinh v_1 k^1, \cosh v_1 k^1, \vec k_T) \;, \nonumber\\
\al q_B^\mu = (V_2 \gamma_2 q^1, \gamma_2 q^1, \vec q_T) = 
(\sinh v_2 q^1, \cosh v_2 q^1, \vec q_T) \;. \nonumber
\eeqa 
Notice that because of their non zero velocity, the branes can also transfer
energy, and not only momentum as in the static case.

Integrating over the bosonic zero modes and taking into account momentum 
conservation ($k_B^\mu = q_B^\mu$), the amplitude factorizes into a bosonic 
and a fermionic partition functions:
\beq
{\cal A}=\frac 1{\sinh v} \int_{0}^{\infty}dl 
\int \frac {d^2 \vec k_T}{(2\pi)^2} e^{i \vec k \cdot \vec b} 
e^{-\frac {q_B^2}2} \sum_s Z_B Z^s_F 
=\frac 1{\sinh v} \int_{0}^{\infty} \frac {dl}{2\pi l} e^{- \frac {b^2}{2l}}
\sum_s Z_B Z^s_F \nn 
\eeq
with
$$
Z_{B,F}=<B,V_1|e^{-lH}|B,V_2>^s_{B,F} \;.
$$

From now on, $X^\mu \equiv X^\mu_{osc}$; moreover, it will prove convenient 
to group the fields into pairs
\beqa
X^\pm = X^0 \pm X^1 \a\rightarrow\a \alpha_{n},\beta_{n}= 
a^{0}_{n} \pm a^{1}_{n} \;, \nonumber \\
X^{i},X^{i*} = X^i \pm i X^{i+1} \a\rightarrow\a 
\beta^i_{n},\beta^{i*}_{n}= a^{i}_{n}\pm i a^{i+1}_{n} \;,\;\; i=2,4,6,8  
\;, \nonumber \\
\chi^{A,B} = \psi^0 \pm \psi^1 \a\rightarrow\a \chi^{A,B}_{n}= 
\psi^{0}_{n}\pm\psi^{1}_{n} \;,\nonumber \\
\chi^{i},\chi^{i*}= \psi^i \pm i \psi^{i+1} \a\rightarrow\a \chi^{i}_{n},
\chi^{i*}_{n}= \psi^{i}_{n}\pm i \psi^{i+1}_{n}
\;,\;\; i=2,4,6,8 \;, \nonumber
\eeqa
with the commutation relations $[\alpha_m,\beta_{-n}] = -2 \delta_{mn}$,
$[\beta_m^i,\beta_{-n}^{i*}] = 2 \delta_{mn}$,
$\{\chi^A_m,\chi^B_{-n}\} = -2 \delta_{mn}$ and 
$\{\chi_m^i,\chi_n^{i*}\} = 2 \delta_{mn}$.
In this way, any rotation or boost will reduce to a simple phase 
transformation on the modes.

\subsection{Orbifold construction}

Let us briefly recall the orbifold construction. An orbifold compactification
can be obtained by identifying points in the compact part of space-time which 
are connected by discrete rotations $g = e^{2\pi i \sum_a z_a J_{aa+1}}$ on 
some of the compact pairs $X^a$,$\chi^a$, a=4,6,8. In order to preserve at 
least one supersymmetry, one has to impose the condition $\sum_a z_a = 0$.

We will consider three case: toroidal compactification on $T_6$ and orbifold
compactification on $T_2 \otimes T_4/Z_2$ and $T_6/Z_3$. The construction is
universal, and these three cases can be obtained by explicit choices for the
angles $z_a$: 
\beqa
\mbox{$T_6/Z_3$ ($N=2$ SUSY)} \a :\; \a \mbox{take $z_4,z_6 = \frac 13, 
\frac 23 \;,\;\; z_8 = -z_4 - z_6$} \;, \nn \\
\mbox{$T_2 \otimes T_4/Z_2$ ($N=4$ SUSY)} \a :\;\a \mbox{take 
$z_4= -z_6 = \frac 12 \;,\;\; z_8 = 0$} \;, \nn \\
\mbox{$T_6$ ($N=8$ SUSY)} \a :\;\a \mbox{take $z_4=z_6 = z_8 = 0$} \;. \nn
\eeqa

The spectrum of the theory now contains additional twisted sectors, in which 
periodicity is achieved only up to an element of the quotient group $Z_N$. 
One can diagonalize the fields such that they satisfy the periodicity condition
($g_a = e^{2\pi i z_a}$)  
$$
X^{a}(\sigma+1)=g_a X^a(\sigma) \;,\;\;
X^{*a}(\sigma+1)=g_a^* X^{*a}(\sigma) \;,
$$
and similarly for fermions. This leads to fractional moding in the compact 
directions.

These twisted states exist at fixed points of the orbifold. 
They thus occur only for the 0-brane of type IIA, which corresponds to 
Dirichlet b.c. in all the compact directions and can thus be localized at 
a fixed point.  

Finally, in all sectors, one has to project onto invariant states to get the 
physical spectrum of the theory which is invariant under orbifold rotations. 
In particular, the physical boundary state is given by the projection
$$
|B_{phys}>=\frac{1}{N} (|B,1>+|B,g>+...+|B,g^{N-1}>)
$$
in terms of the twisted boundary states $|B,g^k> = g^k|B>$.

\subsection{0-brane: untwisted sector}

Consider first the static case. The b.c. are Neumann for time and Dirichlet 
for all other directions (i=2,4,6,8 and a=2,4,6).

For the bosons, the b.c. translate into the following equations
\beqa
\al (\alpha_n+\tilde{\beta}_{-n})|B>_B=0 \;,\;\;
(\beta_n+\tilde{\alpha}_{-n})|B>_B=0 \;, \nn \\
\al (\beta^i_n - \tilde\beta^{i}_{-n} )|B>_B=0 \;,\;\;
(\beta^{i*}_n - \tilde\beta^{i*}_{-n} )|B>_B=0 \;, \nn
\eeqa
The boundary state which solves them is given by a Bogolubov transformation
$$
|B>_B= \exp \left\{\frac{1}{2}\sum_{n = 1}^\infty
(\alpha_{-n} \tilde \alpha_{-n} + \beta_{-n} \tilde \beta_{-n} +
\beta^i_{-n}\tilde\beta^{i*}_{-n} + \beta^{i*}_{-n}\tilde\beta^{i}_{-n})
\right\}|0>\;.
$$

For the fermions, one has integer or half-integer moding in the RR and NSNS
sectors respectively. The b.c lead to
\beqa
\al (\chi^A_n+i\eta\tilde{\chi}^B_{-n})|B,\eta>_F=0 \;,\;\;
(\chi^B_n+i\eta\tilde{\chi}^A_{-n})|B,\eta>_F=0 \;, \nn \\
\al (\chi^i_n -i\eta \tilde\chi^{i}_{-n} )|B,\eta>_F=0 \;,\;\;
(\chi^{i*}_n -i\eta \tilde\chi^{i*}_{-n} )|B,\eta>_F=0 \nn \;.
\eeqa
Here $\eta = \pm 1$ has been introduced to deal later on with the GSO 
projection. 

The corresponding boundary state can be factorized into zero mode and 
oscillator parts:
$$
|B,\eta>_F = |B_{o}>_F \otimes |B_{osc}>_F \;.
$$
The oscillator part is the same for both sectors, with appropriate moding
$$
|B_{osc},\eta>_F=\exp \left\{\frac{i\eta}{2}\sum_{n > 0}
(\chi^A_{-n} \tilde \chi^A_{-n} + \chi^B_{-n} \tilde \chi^B_{-n} -
\chi^i_{-n}\tilde\chi^{i*}_{-n} - \chi^{i*}_{-n}\tilde\chi^{i}_{-n})
\right\}|0> \;.
$$
The zero mode part exists only in the RR sector, and is slightly more subtle
to construct.

Since they satisfy a Clifford algebra, the zero modes are proportional to 
$\Gamma$-matrices $\psi^\mu_o = i/\sqrt{2} \Gamma^\mu$,
$\tilde \psi^\mu_o = i/\sqrt{2} \tilde \Gamma^\mu$.
One can then construct the creation-annihilation operators
$a,a^* = 1/2 (\Gamma^0 \pm \Gamma^1)$, 
$b^i,b^{i*} = 1/2 (-i\Gamma^i \pm \Gamma^{i+1})$
and similarly for tilded operators, satisfying the usual algebra
$\{a,a^*\} = \{b^i,b^{i*}\} = 1$.

The b.c. for the zero modes can then be rewritten as
\beqa
\al (a+i\eta\tilde a^*)|B_{o},\eta>_F=0 \;,\;\;
(a^*+i\eta\tilde a)|B_{o},\eta>_F=0 \;, \nn \\
\al (b^i -i\eta \tilde b^{i})|B_{o},\eta>_F=0 \;,\;\;
(b^{i*} -i\eta \tilde b^{i*})|B_{o},\eta>_F=0 \;, \nn
\eeqa
Defining the spinor vacuum $|0> \otimes |\tilde{0}>$ such that
$a|0> = \tilde a |\tilde{0}> = b^i|0> = \tilde b^{i*} |\tilde{0}> = 0$
the zero mode part of the boundary state can then be written as
$$
|B_o,\eta>_{RR} = \exp \left\{-i\eta (a^* \tilde a^* - b^{i*} \tilde b^i) 
\right\}|0> \otimes |\tilde{0}> \;.
$$

The complete boundary state is already invariant under orbifold rotations, 
for which
\beqa
\label{orb}
\al \beta _n^a \rightarrow g_a \beta_n^a \;,\;\;
\chi_n^a \rightarrow g_a \chi_n^a \;,\;\;
b^a \rightarrow g_a b^a \;, \nn \\
\al \beta_n^{a*} \rightarrow g^*_a \beta_n^{a*} \;,\;\;
\chi_n^{a*} \rightarrow g^*_a \chi_n^{a*} \;,\;\;
b^{a*} \rightarrow g^*_a b^{a*} \;.
\eeqa
This comes from the fact that the $Z_N$ action rotates pairs of fields with 
the same b.c. and is thus irrelevant.

For a boost of rapidity $v$, the transformations on the modes are
\beqa
\label{boost}
\al \alpha_n \rightarrow e^{-v} \alpha_n \;,\;\;
\chi^A_n \rightarrow e^{-v} \chi_n^A \;,\;\;
a \rightarrow e^{-v} a \;, \nn \\
\al \beta_n \rightarrow e^{v} \beta_n \;,\;\;
\chi_n^B \rightarrow e^{v} \chi_n^B \;,\;\;
a^* \rightarrow e^{v} a^* \;.
\eeqa
The spinor vacuum is no longer invariant, but transforms as
$|0> \otimes |\tilde{0}> \rightarrow e^{-v} |0> \otimes |\tilde{0}>$.
Finally, the complete boosted boundary state is
\beqa
\label{bs0}
\al |B,V>_B=\exp \left\{\frac{1}{2}\sum_{n > 0}
(e^{-2v}\alpha_{-n} \tilde \alpha_{-n} + e^{2v} \beta_{-n} \tilde \beta_{-n} +
\beta^i_{-n}\tilde\beta^{i*}_{-n} + \beta^{i*}_{-n}\tilde\beta^{i}_{-n})
\right\}|0> \;, \nn \\
\al |B_{osc},V,\eta>_F= \exp \left\{\frac{i\eta}{2}\sum_{n > 0}
(e^{-2v} \chi^A_{-n} \tilde \chi^A_{-n} + e^{2v} \chi^B_{-n} \tilde 
\chi^B_{-n} - \chi^i_{-n}\tilde\chi^{i*}_{-n} - 
\chi^{i*}_{-n}\tilde\chi^{i}_{-n})\right\}|0> \;, \\
\al |B_o,V,\eta>_{RR} = e^{-v} \exp \left\{
-i\eta (e^{2v} a^* \tilde a^* - b^{i*} 
\tilde b^i) \right\}|0> \otimes |\tilde{0}> \;. \nn
\eeqa

In both sectors, the fermion number operator reverses the sign of the parameter
$\eta$, that is $(-1)^F|B,V,\eta> = - |B,V,-\eta>$, and the GSO-projected 
boundary state is
$$
|B,V> = \frac 12 (|B,V,+>-|B,V,->) \;.
$$

The partition function can then be computed carrying out some simple oscillator
algebra; the ghosts cancel one untwisted pair, say 2-3, and the result is the 
product of the contributions of the 0-1 pair and the 3 compact pairs.

For the bosons, one finds ($q=e^{-2\pi l}$)
\beqa
\al <B,V_1|e^{-lH}|B,V_2>_B^{(0,1)} = \prod_{n=1}^\infty 
\frac 1{(1 - e^{-2v}q^{2n}) (1 - e^{2v}q^{2n})} \;, \nn \\ 
\al <B,V_1|e^{-lH}|B,V_2>_B^{(a,a+1)} = \prod_{n=1}^\infty 
\frac 1{(1 - q^{2n})^2} \;. \nn 
\eeqa
The total bosonic partition function is thus (zero-point energy 
$q^{-\frac 23}$)
\beq
Z_B = 16 \pi^3 i \sinh v q^{\frac 13} f(q^2)^{4} 
\frac 1{\vartheta_1(i\frac v\pi|2il) \vartheta_1^\prime(0|2il)^3} \;.
\eeq

For the fermions, the 0-1 pair gives
$$
<B,V_1,\eta|e^{-lH}|B,V_2,\eta^\prime>_F^{s(0,1)} = Z^s_o(\eta\eta^\prime)
\prod_{n>0} (1 +\eta\eta^\prime e^{-2v}q^{2n})
(1 +\eta\eta^\prime e^{2v}q^{2n}) \;,
$$
with $\eta\eta^\prime=\pm 1$ and the zero mode contributions
$$
Z_o^{R}(+) = 2 \cosh v \;,\;\; Z_o^{R}(-) = 2 \sinh v \;,\;\;
Z_o^{NS}(\pm) = 1 \;. \nn
$$
Each compact pair gives instead
$$
<B,V_1,\eta|e^{-lH}|B,V_2,\eta^\prime>_F^{s(a,a+1)} = Z^s_o(\eta\eta^\prime)
\prod_{n>0} (1 +\eta\eta^\prime q^{2n})^2 \;,
$$
with 
$$
Z_o^{R}(+) = 2 \;,\;\; Z_o^{R}(-) = 0 \;,\;\;
Z_o^{NS}(\pm) = 1 \;. \nn
$$
After the GSO projection, only the three even spin structures R+ and 
NS$\pm$ contribute, and (zero-point energy $q^{- \frac 13}$ for NSNS 
and $q^{\frac 23}$ for RR)
\beqa
\al Z_F=q^{-\frac 13} f(q^2)^{-4}
\left\{\vartheta_2(i\frac{v}{\pi}|2il)\vartheta_2(0|2il)^3
-\vartheta_3(i\frac{v}{\pi}|2il)\vartheta_3(0|2il)^3
+\vartheta_4(i\frac{v}{\pi}|2il)\vartheta_4(0|2il)^3\right\} \nn \\
\al \quad \;\;\sim V^4 \;,
\eeqa
corresponding to the usual cancellation of the force between two BPS states
\cite{Polch,Bachas}.
Thus, the untwisted sector for the 0-brane gives the same result as
the uncompactified theory for every compactification scheme.

\subsection{0-brane: twisted sector}

Consider now the twisted sector, which has to be included when the 0-brane
is at an orbifold fixed point. In this case, the boundary state is similar 
to the one of the untwisted sector, with fractional moding in the compact 
directions.

In the $Z_3$ case, each pair of compact bosons gives
$$
<B,V_1|e^{-lH}|B,V_2>_B^{(a,a+1)} = \prod_{n=1}^\infty 
\frac 1{(1 - q^{2(n-\frac 13)})(1 - q^{2(n-\frac 23)})} \;.
$$
For a pair of compact fermions (no zero modes)
\beqa
\al <B,V_1,\eta|e^{-lH}|B,V_2,\eta^\prime>_{R}^{s(a,a+1)} = 
\prod_{n=1}^\infty (1 +\eta\eta^\prime q^{2(n-\frac 13)})
(1 +\eta\eta^\prime q^{2(n-\frac 23)}) \;, \nn \\
\al <B,V_1,\eta|e^{-lH}|B,V_2,\eta^\prime>_{NS}^{s(a,a+1)} =
\prod_{n=1}^\infty (1 +\eta\eta^\prime q^{2(n-\frac 16)})
(1 +\eta\eta^\prime q^{2(n-\frac 56)}) \;. \nn
\eeqa
The total partition functions after the GSO projection are 
(the zero-point energies add to zero)
\beqa
\al Z_B=2 i \sinh v f(q^2)^4 \frac 1{\vartheta_1(i \frac v\pi|2il)
\vartheta_1(-\frac 23 il|2il)^3} \;, \\
\al Z_F=f(q^2)^{-4}
\left\{\vartheta_2(i\frac{v}{\pi}|2il)\vartheta_2(-\frac 23 il|2il)^3
-\vartheta_3(i\frac{v}{\pi}|2il) \vartheta_3(-\frac 23 il|2il)^3
-\vartheta_4(i\frac{v}{\pi}|2il)\vartheta_4(-\frac 23 il|2il)^3\right\} \nn \\
\al\quad \;\;\sim V^2 \;.
\eeqa

In the $Z_2$ case, the analysis is similar and the results are
\beqa
\al Z_B=2 i \sinh v q^{-\frac 16} f(q^2)^4 
\frac 1{\vartheta_1(i \frac v\pi|2il)
\vartheta_1(0|2il) \vartheta_1(-il|2il)^2} \;, \\
\al Z_F=q^{\frac 16} f(q^2)^{-4}
\left\{\vartheta_2(i\frac{v}{\pi}|2il)\vartheta_2(0|2il)
\vartheta_2(-il|2il)^2 \right. 
\nn \\ \al \qquad \qquad \qquad \quad \;\; 
\left. -\vartheta_3(i\frac{v}{\pi}|2il)\vartheta_3(0|2il)
\vartheta_3(-il|2il)^2- 
\vartheta_4(i\frac{v}{\pi}|2il)\vartheta_4(0|2il)
\vartheta_4(-il|2il)^2\right\} \nn \\
\al \quad \;\;\sim V^2 \;.
\eeqa

\subsection{3-brane}

Let us now consider a particular 3-brane configuration. In the static case, 
we take Neumann b.c. for time, Dirichlet b.c. for space and mixed b.c. for 
each pair of compact directions, say Neumann for the a directions and 
Dirichlet for the a+1 directions.

The new b.c. for the compact directions are
\beqa
\al (\beta^a_n + \tilde\beta^{a*}_{-n} )|B>_B=0 \;,\;\;
(\beta^{a*}_n + \tilde\beta^{a}_{-n} )|B>_B=0 \;, \nn \\
\al (\chi^a_n +i\eta \tilde\chi^{a*}_{-n} )|B_{osc},\eta>_F=0 \;,\;\;
(\chi^{a*}_n +i\eta \tilde\chi^{a}_{-n} )|B_{osc},\eta>_F=0 \;, \nn \\
\al (b^a +i\eta \tilde b^{a*})|B_{o},\eta>_F=0 \;,\;\;
(b^{a*} +i\eta \tilde b^{a})|B_{o},\eta>_F=0 \;. \nn
\eeqa
Defining a new spinor vacuum $|0> \otimes |\tilde{0}>$ such that
$b^a|0> = \tilde b^{a} |\tilde{0}> = 0$ the compact part of the boundary 
state is
\beqa
\al |B>_B=\exp \left\{-\frac{1}{2}\sum_{n > 0}
(\beta^a_{-n}\tilde\beta^{a}_{-n} 
+ \beta^{a*}_{-n}\tilde\beta^{a*}_{-n})\right\}|0> \;, \nn \\
\al |B_{osc},\eta>_F=\exp \left\{\frac{i\eta}{2}\sum_{n > 0}
(\chi^a_{-n}\tilde\chi^{a}_{-n} + \chi^{a*}_{-n}\tilde\chi^{a*}_{-n})
\right\}|0> \;, \nn \\
\al |B_o,\eta>_{RR} = \exp \left\{-i\eta b^{a*} \tilde b^{a*} \right\} 
|0> \otimes |\tilde{0}> \;. \nn
\eeqa

In this case, the boundary state is not invariant under orbifold 
rotations, under which the modes of the fields transform as in 
eq. (\ref{orb}) and the spinor vacuum as 
$|0> \otimes |\tilde{0}> \rightarrow g_a |0> \otimes |\tilde{0}>$.
This was expected since a $Z_N$ rotation now mixes two directions with 
different b.c, and thus the corresponding closed string state does not 
need to be invariant under $Z_N$ rotations.

The compact part of the twisted boundary state is finally found to be
\beqa
\label{bs3}
\al |B,V,g_a>_B= \exp \left\{-\frac{1}{2}\sum_{n > 0}
(g_a^2 \beta^a_{-n}\tilde\beta^{a}_{-n} + 
g_a^{*2}\beta^{a*}_{-n}\tilde\beta^{a*}_{-n})\right\}|0> \;, \nn \\
\al |B_{osc},V,g_a,\eta>_F=\exp \left\{\frac{i\eta}{2}\sum_{n > 0}
(g_a^{2} \chi^a_{-n}\tilde\chi^{a}_{-n} 
+ g_a^{*2} \chi^{a*}_{-n}\tilde\chi^{a*}_{-n})\right\}|0> \;, \\
\al |B_o,V,g_a,\eta>_{RR} = g_a \exp \left\{-i\eta g_a^{*2} b^{a*} 
\tilde b^{a*} \right\}|0> \otimes |\tilde{0}> \;. \nn
\eeqa

Each pair of compact bosons gives now a contribution to the partition
function which depends on the orbifold relative angle 
($(g_a^* g_a^\prime)^2 = e^{2\pi i w_a}$)
$$
<B,V_1,g_a|e^{-lH}|B,V_2,g_a^\prime>_B^{(a,a+1)} = 
\prod_{n=1}^\infty \left|\frac 1{1 +\eta\eta^\prime e^{2\pi i w_a}q^{2n}}
\right|^2 \;.
$$
For fermions one obtains
$$
<B,V_1,g_a,\eta|e^{-lH}|B,V_2,g_a^\prime,\eta^\prime>_F^{s(a,a+1)} = 
Z^s_o(\eta\eta^\prime) 
\prod_{n>0} \left|1 +\eta\eta^\prime e^{2\pi i w_a}q^{2n}\right|^2 \;,
$$
where
$$
Z_o^{R}(+) = 2 \cos \pi w_a \;,\;\; Z_o^{R}(-) = 2 i \sin \pi w_a \;,\;\;
Z_o^{NS}(\pm) = 1 \;.
$$

After the GSO projection, the total partition functions for a given relative
angle $w_a$ are
\beq
Z_B=16 i \sinh v q^{\frac 13} f(q^2)^4 
\frac 1{\vartheta_1(i \frac v\pi|2il)} 
\prod_a \frac {\sin \pi w_a}{\vartheta_1(w_a|2il)} \;,
\eeq
\beqa
\al Z_F=q^{-\frac 13}f(q^2)^{-4}
\left\{\vartheta_2(i\frac{v}{\pi}|2il)\prod_a \vartheta_2(w_a|2il) 
\right. \nn \\ \al \qquad \qquad \qquad \qquad \; \left.
-\vartheta_3(i\frac{v}{\pi}|2il)\prod_a \vartheta_3(w_a|2il)
+\vartheta_4(i\frac{v}{\pi}|2il)\prod_a \vartheta_4(w_a|2il)\right\} \nn \\
\al  \quad \;\; \sim \left\{
\begin{array}{l}
V^4 \;\;,\;\;w_a=0 \\
V^2 \;\;,\;\;w_a \neq 0 
\end{array}
\right. \;.
\eeqa
Recall that to obtain the invariant amplitude, one has to average over all 
possible angles $w_a$. 

Finally, for this 3-brane configuration there is no twisted sector, as already
explained.

\subsection{Large distance limit and field theory interpretation}

In the large distance limit $l \rightarrow \infty$, explicit results with
their exact dependence on the rapidity can be obtained and compared to 
a field theory computation. The behaviors that one finds are the following:

\noindent
{\bf 0-brane}

\noindent
a) Untwisted sector
\beq
{\cal A} \sim 4 \cosh v - \cosh 2v - 3 \sim V^4 \;.
\eeq
b) Twisted sector
\beq
{\cal A} \sim \cosh v - 1 \sim V^2 \;.
\eeq

\noindent
{\bf 3-brane}
\beqa
\al {\cal A}(w_a) \sim 4 \prod_a \cos \pi w_a \cosh v - \cosh 2v - 
\sum_a \cos 2 \pi w_a \;, \nn \\
\al {\cal A} \sim \left\{
\begin{array}{l}
\cosh v - \cosh 2v \sim V^2 \;\;,\;\; T_6/Z_3 \\
4 \cosh v - \cosh 2v - 3 \sim V^4\;\;,\;\; T_2 \otimes T_4/Z_2 \;,\; T_6
\end{array}
\right. \;.
\eeqa

In the low energy effective supergravity field theories, the possible 
contributions to the scattering amplitude in the eikonal approximation come
from vector exchange in the RR sector and dilaton and graviton exchange in 
the NSNS sector. 
The respective contributions have a peculiar dependence on the 
rapidity reflecting the tensorial nature and are: 
\beq
{\cal A}^{NS}_{\phi} \sim -a^2 \;,\;\;
{\cal A}^{R}_{V_\mu} \sim e^2 \cosh v \;,\;\;
{\cal A}^{NS}_{g_{\mu\nu}} \sim -M^2 \cosh 2v \;.
\eeq

Thus, the interpretation of the behaviors found in the various sectors
and for the various brane configurations we have considered, is the following:
\beqa
4 \cosh v - \cosh 2v - 3 \quad \a \Leftrightarrow \a \quad 
\mbox{$N=8$ Grav. multiplet} \;, \nn \\
\cosh v - \cosh 2v \quad \a \Leftrightarrow \a \quad
\mbox{$N=2$ Grav. multiplet} \;, \nn \\
\cosh v - 1 \quad \a \Leftrightarrow \a \quad \mbox{Vec. multiplet} \;. \nn
\eeqa

The patterns of cancellation suggest that all the D-brane configurations
that we have considered correspond to  extremal p-brane solutions of the 
low energy supergravity, possibly coupling to the additional twisted vector 
multiplets; the 3-brane configuration on the $Z_3$ orbifold seems to be an
exception since it does not couple to the scalars, and should thus correspond
to a Reissner-Nordstr\"om extremal black hole. 

Finally, notice that $V^2$ terms in the effective action give a non flat
metric to the moduli space. Since in the dual open string channel a constant
velocity $V$ corresponds by $T$-duality to a constant electric field $E$, 
$V^2$ terms correspond to a renormalization of the Maxwell term $E^2$.
It is well known that this can not happen for maximally supersymmetric 
theories; the $V^2$ behavior is thus forbidden for $N=8$ compactifications,
but generically allowed for compactifications breaking some supersymmetry,
$N < 8$. Our results are compatible with this and show that $V^2$ terms do 
indeed appear in some cases.

\section{Emission of massless NSNS bosons}
 
Consider two moving D-branes in interaction emitting a massless NSNS boson.

\begin{figure}[h]
\centerline{\psfig{figure=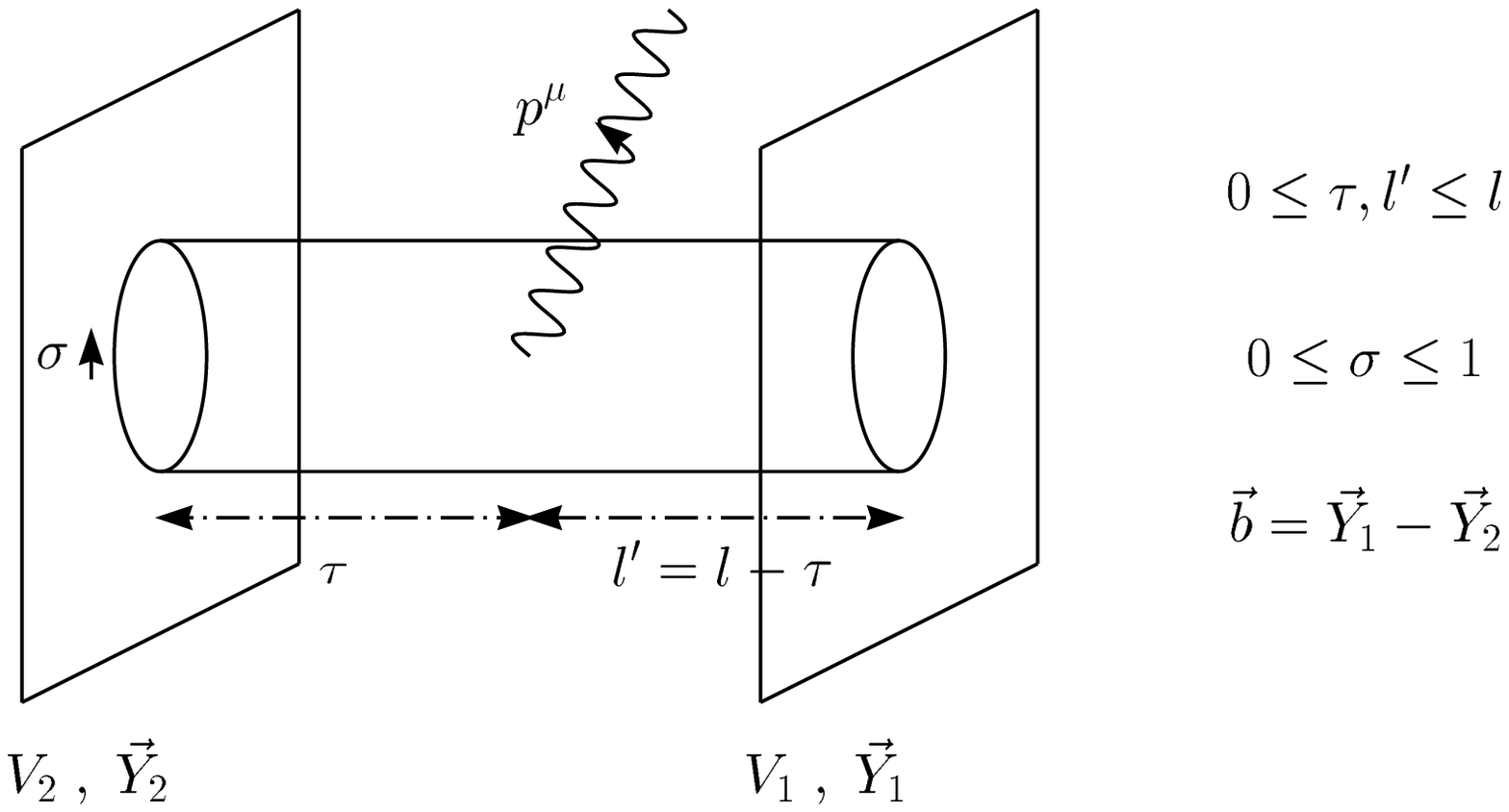,width=300pt}}
\label{fig2}
\end{figure}

The amplitude is computed inserting the usual vertex operator 
($z=\sigma + i \tau$)
$$
V(z, \bar z) = 
G_{ij} (\partial X^i - {1\over 2}p\cdot \psi \psi^i) 
(\bar \partial X^j + {1\over 2}p \cdot \bar\psi \bar \psi^j) 
e^{i p \cdot X}
$$
between the two boundary states
\beqa
\al {\cal A}=\int_0^\infty dl \int_0^l d\tau \sum_s
<B,V_1,\vec Y_1| e^{-lH} V(z, \bar z) |B,V_2,\vec Y_2>_s \nn \\
\al \quad \,= \int_0^\infty d\tau \int_0^\infty dl^\prime \sum_s 
<V(z, \bar z)>_s \;. \nn
\eeqa
We have chosen a purely space-like polarization tensor as allowed by gauge
invariance.

As before, we split the bosons into zero mode and oscillators to be treated 
separately (again $X^\mu \equiv X^\mu_{osc}$).
As usual, the zero mode part ensures momentum conservation 
($p^\mu = k_B^\mu - q_B^\mu$) and gives the kinematics.
The energies and longitudinal momenta are completely fixed
by the momentum of the outgoing particle
($\cos \theta = p^1/p$, $p=p^0$),
\begin{eqnarray}
\al k_B^0 = V_1 k_B^1 \;,\;\; 
k_B^1 = \frac p{V_1 - V_2} (1 - V_2 \cos \theta) \;, \nn \\
\al q_B^0 = V_2 q_B^1  \;,\;\;
q_B^1 = \frac p{V_1 - V_2} (1 - V_1 \cos \theta) \;. \nonumber
\end{eqnarray}

The zero mode contribution is found to have the simple structure 
($v = v_1 - v_2$)
$$
<e^{i p \cdot X}>_o = \frac 1{\sinh v}  
\int \frac {d^2 \vec k_{T}}{(2 \pi)^2} e^{i \vec k \cdot \vec b} 
e^{- \frac {q^2}2 \tau} e^{- \frac {k^2}2 l^\prime} \;.
$$
Further zero mode insertions give just additional momentum factors
$$
\partial X^i_o \;\Rightarrow\; -\frac 12 k_B^i \;,\;\;
\bar \partial X^j_o \;\Rightarrow\;  \frac 12 k_B^j \;,\;\;
\partial X^i_o \bar \partial X^j_o \;\Rightarrow\;  -\frac 14 
k_B^i k_B^j \;. \nn
$$ 

Finally, the amplitude can be rewritten (from now on $q^\mu \equiv 
q_B^\mu$ and $k^\mu \equiv k_B^\mu$) as
\beq
{\cal A} = \frac 1{\sinh v} \int_0^\infty d\tau 
\int_0^\infty dl^\prime \int \frac {d^2 \vec k_{T}}{(2 \pi)^2} 
e^{i \vec k\cdot \vec b}e^{- \frac {q^2}2 \tau} 
e^{- \frac {k^2}2 l^\prime} <e^{i p \cdot X}>
\sum_s Z_B Z_F^{s}{\cal M}_s \;,
\eeq
with
\begin{eqnarray}
\label{contr}
\al {\cal M}^s = G_{ij} \left\{<\partial X^i \bar \partial X^j>
- <\partial X^i p \cdot X> <\bar \partial X^j p \cdot X> 
\right. \nonumber \\
\al
\qquad \qquad \;\;\; +\frac 14 \left(<p \cdot \psi p \cdot \bar \psi>_s 
<\psi^i \bar \psi^j>_s -<p \cdot \psi \psi^i>_s 
<p \cdot \bar \psi \bar \psi^j>_s \right. \nonumber \\ 
\al \qquad \qquad \qquad \;\;\;\left. + 
<p \cdot \bar \psi \psi^i>_s <p \cdot \psi \bar \psi^j>_s
\right) \nonumber \\
\al \qquad \qquad \;\;\; + \frac i2 \left(<\partial X^i p \cdot X> 
<p \cdot \bar \psi \bar \psi^j>_s
- <\bar \partial X^j p \cdot X> <p \cdot \psi \psi^i>_s \right) 
\nonumber \\
\al \qquad \qquad \;\;\; - \frac 12 k^i \left(i<\bar \partial X^j 
p \cdot X> + \frac 12 <p \cdot \bar \psi \bar \psi^j>_s \right) 
%\nonumber \\ \al \qquad \qquad \;\;\; 
+ \frac 12 k^j \left(i<\partial X^i 
p \cdot X> - \frac 12 <p \cdot \psi \psi^i>_s \right) \nn \\
\al \qquad \qquad \;\;\,\left. - \frac 14 k^i k^j \right\} \;.
\end{eqnarray}
Obviously, the partition function factorizes, leaving connected correlators. 
In the odd spin structure, appropriate zero mode insertion is understood in
order for these expressions to make sense. 

\subsection{Correlators}

The boundary state formalism provides a systematic way of computing 
correlators with non trivial b.c., such as those needed here, through the 
definitions
\beqa
\al <X^\mu X^\nu>=
\frac{<B_1,V_1|e^{-lH}X^\mu X^\nu|B_2,V_2>_B}
{<B_1,V_1|e^{-lH}|B_2,V_2>_B} \;, \\
\al <\psi^\mu \psi^\nu>_s =
\frac{<B_1,V_1,\eta|e^{-lH}\psi^\mu \psi^\nu|B_2,V_2,\eta^\prime>_F^s}
{<B_1,V_1,\eta|e^{-lH}|B_2,V_2,\eta^\prime>_F^s} \;.
\eeqa

For the bosons, one obtains an infinite series of logarithms corresponding 
to the propagation of all the string states with growing mass 
($q=e^{-2\pi\tau}$):
\beqa
\al <X^0(z) \bar X^0(\bar z)>=<X^1(z) \bar X^1(\bar z)>= \nn \\
\al \qquad = \frac 1{4\pi} \sum_{n=0}^\infty
\left\{\cosh 2[(v_1-v_2)n - v_2] \ln (1 - q^{2n} e^{-4\pi \tau})\right. 
\nn \\ \al \qquad \qquad \qquad \;\; 
-\left. \cosh 2[(v_2-v_1)n - v_1] \ln (1 - q^{2n} e^{-4\pi l^\prime})
\right\} \;, \nn \\
\al <X^0(z) \bar X^1(\bar z)>=<X^1(z) \bar X^0(\bar z)>= \nn \\
\al \qquad = -\frac 1{4\pi} \sum_{n=0}^\infty
\left\{\sinh 2[(v_1-v_2)n - v_2] \ln (1 - q^{2n} e^{-4\pi \tau})\right. \nn \\
\al \qquad \qquad \qquad \quad \; 
+\left. \sinh 2[(v_2-v_1)n - v_1] \ln (1 - q^{2n} e^{-4\pi l^\prime})
\right\} \;. \nn 
\eeqa
For the fermions in the NS$\pm$ sectors, one has poles instead of logarithms,
with a similar structure 
\beqa
\al<\psi^0(z) \bar \psi^0(\bar z)>_{NS\pm}=
<\psi^1(z) \bar \psi^1(\bar z)>_{NS\pm}= \nn \\
\al \qquad = -i\sum_{n=0}^\infty (\mp)^n
\left\{\cosh 2[(v_1-v_2)n - v_2] 
\frac {q^{n} e^{-2\pi \tau}}{1 - q^{2n} e^{-4\pi \tau}}\right. \nn \\
\al \qquad \qquad \qquad \qquad \;\; 
\pm \left. \cosh 2[(v_2-v_1)n - v_1] 
\frac {q^{n} e^{-2\pi l^\prime}}{1 - q^{2n} e^{-4\pi l^\prime}} \right\} 
\;, \nn \\ 
\al <\psi^0(z) \bar \psi^1(\bar z)>_{NS\pm}=
<\psi^1(z) \bar \psi^0(\bar z)>_{NS\pm}= \nn \\
\al \qquad = i\sum_{n=0}^\infty (\mp)^n
\left\{\sinh 2[(v_1-v_2)n - v_2] 
\frac {q^{n} e^{-2\pi \tau}}{1 - q^{2n} e^{-4\pi \tau}} \right. \nn \\
\al \qquad \qquad \qquad \qquad \; 
\pm\left. \sinh 2[(v_2-v_1)n - v_1] 
\frac {q^{n} e^{-2\pi l^\prime}}{1 - q^{2n} e^{-4\pi l^\prime}} \right\} 
\;,\nn 
\eeqa
For the fermions in the R$\pm$ sectors, the results are similar,
\beqa
\al <\psi^0(z) \bar \psi^0(\bar z)>_{R\pm}=
<\psi^1(z) \bar \psi^1(\bar z)>_{R\pm}= \nn \\
\al \qquad = F_o^{R}(\pm) -i\sum_{n=0}^\infty (\mp)^n
\left\{\cosh 2[(v_1-v_2)n - v_2] 
\frac {q^{2n} e^{-4\pi \tau}}{1 - q^{2n} e^{-4\pi \tau}}\right. \nn \\
\al \qquad \qquad \qquad \qquad \qquad \qquad \;\; 
\pm \left. \cosh 2[(v_2-v_1)n - v_1] 
\frac {q^{2n} e^{-4\pi l^\prime}}{1 - q^{2n} e^{-4\pi l^\prime}} \right\} 
\;, \nn \\
\al <\psi^0(z) \bar \psi^1(\bar z)>_{R\pm}=
<\psi^1(z) \bar \psi^0(\bar z)>_{R\pm}= \nn \\
\al \qquad = G_o^{R}(\pm)+i\sum_{n=0}^\infty (\mp)^n
\left\{\sinh 2[(v_1-v_2)n - v_2] 
\frac {q^{2n} e^{-4\pi \tau}}{1 - q^{2n} e^{-4\pi \tau}} \right. \nn \\
\al \qquad \qquad \qquad \qquad \qquad \qquad \;\; 
\pm\left. \sinh 2[(v_2-v_1)n - v_1] 
\frac {q^{2n} e^{-4\pi l^\prime}}{1 - q^{2n} e^{-4\pi l^\prime}} \right\} 
\;, \nn
\eeqa
with additional zero mode contributions,
\beqa
\al \qquad F_o^{R}(+) = - \frac i2 \frac {\cosh (v_1 + v_2)}{\cosh (v_1 - v_2)}
\;,\;\; F_o^{R}(-) = - \frac i2 \frac {\sinh (v_1 + v_2)}
{\sinh (v_1 - v_2)} \;, \nn \\
\al \qquad G_o^{R}(+) = - \frac i2 \frac {\sinh (v_1 + v_2)}
{\cosh (v_1 - v_2)} \;,\;\; 
G_o^{R}(-) = - \frac i2 \frac {\cosh (v_1 + v_2)}
{\sinh (v_1 - v_2)} \;. \nn 
\eeqa

As usual, world-sheet supersymmetry means (here for osc.) a relation 
between the odd fermions and the derivative of the bosons 
\beq
\label{susy}
<\partial X^\mu (z) \bar X^\nu (\bar z)> = \frac 12
<\psi^\mu (z) \bar \psi^\nu (\bar z)>_{R-} \;.
\eeq
There are also non vanishing equal-point correlators, which can be computed 
in the same way. They can also be deduced from the previous ones using
the b.c. to reflect left and right movers at the boundaries.

The correlators can be actually expressed in terms of twisted 
$\vartheta$-functions. To understand this, consider the rescaled combinations 
of fermions $\psi^\pm = e^{\mp v_2} (\psi^0 \pm \psi^1)$,
satisfying usual b.c. on one brane and twisted b.c. on the other brane
\begin{eqnarray} 
\al \psi^\pm(z) = -i \bar \psi^\mp (\bar z)
\;,\;\; \tau = 0 \; \Leftrightarrow \; z = \bar z \;, \nonumber \\
\al \psi^\pm(z) = -i e^{\pm 2 v} \bar \psi^\mp (\bar z)
\;,\;\; \tau = l \; \Leftrightarrow \; z = \bar z + 2il \;. \nn
\end{eqnarray}
The propagators 
$P_{(\pm)}^s(z - \bar z) = <\psi^\pm (z)  \bar \psi^\pm (\bar z)>_s$
should accordingly have appropriate periodicity conditions on the covering
torus with modulus 2il from which the cylinder can be obtained by the
involution $z \doteq \bar z + 2il$.

In fact, under the shift $w \rightarrow w + m +2iln$
on the covering torus, the propagators transform as
\begin{eqnarray}
P_{(\pm)}^{R+}(w + m +2iln) \a=\a 
e^{i \pi n} e^{\pm 2 n v} P_{(\pm)}^{R+}(w) \;, \nonumber\\
P_{(\pm)}^{R-}(w + m +2iln) \a=\a 
e^{\pm 2n v} P_{(\pm)}^{R-}(w) \;, \nonumber\\
P_{(\pm)}^{NS+}(w + m +2iln) \a=\a 
e^{i \pi m} e^{i \pi n} e^{\pm 2nv} P_{(\pm)}^{NS+}(w) \;, \nonumber \\
P_{(\pm)}^{NS-}(w + m +2iln) \a=\a 
e^{i \pi m} e^{\pm 2nv} P_{(\pm)}^{NS-}(w) \;, \nn
\end{eqnarray}

These properties, together with the universal local behavior
$P_{(\pm)}^s(w) \rightarrow 1/(4\pi w)$, imply that for the even 
spin structures
\beq
P_{(\pm)}^s(w) = \frac 1{4\pi}
\frac {\vartheta_s (w \pm i \frac v\pi|2il) \vartheta_1^\prime (0 |2il)}
{\vartheta_s (\pm i \frac v\pi |2il) \vartheta_1 (w |2il)} \;.
\eeq

\subsection{Axion}

For the axion, the polarization tensor is transverse and antisymmetric, and
can be taken to be $G_{ij} = 1/2 \epsilon_{ijk} p^k/p$.
Only the odd spin structure can contribute in this case because of the 
antisymmetry of $G_{ij}$. Notice that in the twisted sector of the $Z_3$ 
orbifold, there are only two fermionic zero modes in the 2-3 pair, and the 
amplitude could be non vanishing since there is the possibility of soaking up
these two zero modes with the vertex operator. 

After integrating by parts the two-derivative bosonic term appearing in the 
contraction (\ref{contr}), and using world-sheet supersymmetry (\ref{susy}), 
the result simplifies to
\beq
{\cal M}^{R-}_{ax} = \frac i8 \cos \theta \left[ 
-\partial_\tau <p \cdot X(z) p \cdot \bar X (\bar z)> 
+ \frac 12 (k^2 - q^2) \right] \;.
\eeq
However, since $\partial_\tau |_{l}=\partial_{\tau} |_{l^\prime} -
\partial_{l^\prime} |_\tau$ the final amplitude is a total derivative
($Z_B Z_F^{R-}=2\sinh v$ for the twisted sector of $Z_3$) and vanishes
even in the twisted sector of the $Z_3$ orbifold
\begin{eqnarray}
\al {\cal A}_{ax}= \frac i4 \cos \theta 
\int_0^\infty d\tau \int_0^\infty dl^\prime 
\int \frac {d^2 \vec k_{T}}{(2 \pi)^2} 
e^{i \vec k \cdot \vec b}(\partial_\tau - \partial_{l^\prime}) 
\left\{e^{- \frac {q^2}2 \tau} e^{- \frac {k^2}2 l^\prime} 
<e^{i p \cdot X}> \right\} \nn \\ \al \qquad =0 \;.
\end{eqnarray}

\subsection{Dilaton}

For the dilaton, the polarization tensor is
$G_{ij} = \delta_{ij} - p^i p^j/p^2$.
Only the even spin structures will now contribute, because of the symmetry of 
$G_{ij}$. Again, the two-derivative bosonic term in the contraction is 
conveniently integrated by parts.

In this case, we shall analyze the large distance limit, in which it is enough
to keep only leading terms for $l \rightarrow \infty$ in the propagators. In
this limit, the bosonic exponential reduces to
\beq
\label{exp}
<e^{i p \cdot X}> = 
\left(1 - e^{-4\pi \tau} \right)^{-\frac {p^{(2)2}}{2\pi}}
\left(1 - e^{-4\pi l^\prime} \right)^{-\frac {p^{(1)2}}{2\pi}} \;,
\eeq
with the boosted energies
$p^{(1,2)} = p \gamma_{1,2} (1 - V_{1,2} \cos \theta)
= p (\cosh v_{1,2} - \sinh v_{1,2} \cos \theta)$.

After some complicated algebra, one finds for the contractions (keeping a 
subleading term in the NS$\pm$ sectors because of a possible enhancement 
coming from the partition function)
\beqa
\al {\cal M}^{R+}_{dil} =
\frac 1{4p^2} \left[(k^2 - q^2) - 2 p^2 \cos \theta \tanh v \right] 
%\times \nn \\ \al \qquad \qquad \quad \;\;
\left\{ \frac 14 (k^2 - q^2) - p^{(2)2} \frac {e^{-4\pi \tau}}
{1 - e^{-4\pi \tau}} + p^{(1)2} \frac {e^{-4\pi l^\prime}}
{1 - e^{-4\pi l^\prime}} \right\} \nn \\
\al \qquad \quad \;\; - \frac {k^0}p \left(\frac {q^2}4 + 
p^{(2)2} \frac {e^{-4\pi \tau}}{1 - e^{-4\pi \tau}} \right)
+ \frac {q^0}p \left(\frac {k^2}4 + 
p^{(1)2} \frac {e^{-4\pi l^\prime}}{1 - e^{-4\pi l^\prime}} \right) \;, \\
\al {\cal M}^{NS\pm}_{dil} =
\frac 1{4p^2} \left[(k^2 - q^2) \mp 8 e^{-2\pi l} 
p^2 \cos \theta \sinh v \right] 
%\times \nn \\ \al \qquad \qquad \qquad \;
\left\{ \frac 14 (k^2 - q^2) - p^{(2)2} \frac {e^{-4\pi \tau}}
{1 - e^{-4\pi \tau}} + p^{(1)2} \frac {e^{-4\pi l^\prime}}
{1 - e^{-4\pi l^\prime}} \right\} \nn \\
\al \qquad \qquad \; - \frac {k^0}p \left(\frac {q^2}4 + 
p^{(2)2} \frac {e^{-4\pi \tau}}{1 - e^{-4\pi \tau}} \right)
+ \frac {q^0}p \left(\frac {k^2}4 + 
p^{(1)2} \frac {e^{-4\pi l^\prime}}{1 - e^{-4\pi l^\prime}} \right) \;.
\eeqa

Using eq. (\ref{exp}) for $<e^{ip \cdot X}>$ and integrating by parts in the 
final amplitude, one finds the following rules for the $\tau$ and $l^\prime$
poles in the contraction:
\beq
\label{equiv}
\frac {e^{-4\pi \tau}}{1 - e^{-4\pi \tau}} \doteq
- \frac 14 \frac {q^2}{p^{(2)2}} \;,\;\;
\frac {e^{-4\pi l^\prime}}{1 - e^{-4\pi l^\prime}} \doteq 
- \frac 14 \frac {k^2}{p^{(1)2}} \;.
\eeq
These imply ${\cal M}^{R+}_{dil}={\cal M}^{NS\pm}_{dil}=0$, and thus at
large distances
\beq
{\cal A}_{dil} = 0 \;.
\eeq

\subsection{Graviton}

For the graviton, the polarization tensor is taken to be symmetric transverse
and traceless, $G_{ij} = h_{ij} = h_{ji} \;\;,\;\; p^i h_{ij} = h^i_i = 0$,
and has two independent components. In this case, things are more 
complicated, but one can proceed essentially as for the dilaton. One obtains 
for $l \rightarrow \infty$
\begin{eqnarray}
\al {\cal M}^{R+}_{grav} = 
-\frac 14 \left[h_{ij}k^i k^j - p \tanh v h_{i1} k^i \right] \nn \\
\al \qquad \qquad \, -V_2 \gamma_2 \left[p^{(2)} \left(h_{i1}k^i 
- \frac p2 \tanh v h_{11}\right) 
+ \frac 14 (k^2 - q^2) V_2 \gamma_2 h_{11} \right]
\frac{e^{-4\pi\tau}}{1 - e^{-4\pi\tau}} \nonumber \\
\al \qquad \qquad \, +V_1 \gamma_1 \left[p^{(1)} \left(h_{i1}k^i - 
\frac p2 \tanh v h_{11}
\right) + \frac 14 (k^2 - q^2) V_1 \gamma_1 h_{11} \right]
\frac{e^{-4\pi l^\prime}}{1 - e^{-4\pi l^\prime}} \;, \\
\al  {\cal M}^{NS\pm}_{grav} =
-\frac 14 \left[h_{ij}k^i k^j \mp 4 e^{-2\pi l}
\left(p \sinh 2v h_{i1}k^i - p^2 \sinh^2 v h_{11} \right) \right] \nn \\
\al \qquad \qquad \; -V_2 \gamma_2 \left[p^{(2)} \left(h_{i1}k^i 
\mp 2 e^{-2\pi l} p \sinh v h_{11} \right) 
+ \frac 14 (k^2 - q^2) V_2 \gamma_2 h_{11} \right]
\frac{e^{-4\pi\tau}}{1 - e^{-4\pi\tau}} \nonumber \\
\al \qquad \qquad \; +V_1 \gamma_1 \left[p^{(1)} \left(h_{i1}k^i
\mp 2 e^{-2\pi l} p \sinh v h_{11}
\right) + \frac 14 (k^2 - q^2) V_1 \gamma_1 h_{11} \right] 
\frac{e^{-4\pi l^\prime}}{1 - e^{-4\pi l^\prime}} \;.
\end{eqnarray}

One can use the same equivalence relations (\ref{equiv}) as before to write 
${\cal M}^s_{grav}$ in a $\tau,l^\prime$-independent form; but in any case,
${\cal A}_{grav} \neq 0$. The general structure of the amplitude is
\beq
{\cal A}_{grav} = \frac 1{\sinh v} 
\int \frac {d^2 \vec k_{T}}{(2 \pi)^2} e^{i \vec k\cdot \vec b}
I_1 I_2 \sum_s Z_B Z_F^{s}{\cal M}^s_{grav} \;,
\eeq
and involves three independent functions of the momenta 
$$
{\cal M}^s_{grav} = B^s (p,k,q) + q^2 C^s_1 (p,k,q) + k^2 C^s_2 (p,k,q) \;. 
$$
The kinematical integrals over the two proper times $\tau,l^\prime$ 
can be easily evaluated, finding the usual dual structure with a double 
serie of poles
\beqa
\al I_1 = \int_0^\infty d\tau e^{- \frac {q^2}2 \tau}
\left(1 - e^{-4\pi \tau} \right)^{-\frac {p^{(2)2}}{2\pi}}
=-\frac{1}{4\pi}\frac{\Gamma [\frac{q^2}{8\pi}]\Gamma [-\frac{p^{(2)2}}
{2\pi}+1]}{\Gamma [\frac{q^2}{8\pi}-\frac{p^{(2)2}}{2\pi}+1]} 
\hspace{4pt}\longrightarrow\hspace{-20pt}\raisebox{-6pt}
{$\scriptscriptstyle{p \rightarrow 0}$}\; -\frac 2{q^2} \;, \nn \\
\al I_2 = \int_0^\infty d l^\prime e^{- \frac {k^2}2 l^\prime}
\left(1 - e^{-4\pi l^\prime} \right)^{-\frac {p^{(1)2}}{2\pi}}
=-\frac{1}{4\pi}\frac{\Gamma [\frac{k^2}{8\pi}]\Gamma [-\frac{p^{(1)2}}
{2\pi}+1]}{\Gamma [\frac{k^2}{8\pi}-\frac{p^{(1)2}}{2\pi}+1]} 
\hspace{4pt}\longrightarrow\hspace{-20pt}\raisebox{-6pt}
{$\scriptscriptstyle{p \rightarrow 0}$}\; -\frac 2{k^2} \;.\nn
\eeqa
The last limit is required by the eikonal approximation ($p \ll M=1$) and 
selects the massless part of the states emitted by the branes.

Finally, the amplitude assumes a simple field theory form
\beq
{\cal A}_{grav} = \frac 4{\sinh v} 
\int \frac {d^2 \vec k_{T}}{(2 \pi)^2} e^{i \vec k\cdot \vec b}
\left\{ B^s \frac 1{q^2 k^2} + C^s_1 \frac 1{k^2} 
+ C^s_2 \frac 1{q^2} \right\} \;.
\eeq
The graphical interpretation of the three contributions $B^s$, $C_1^s$ and 
$C_2^s$ is the following

\begin{figure}[h]
\centerline{\psfig{figure=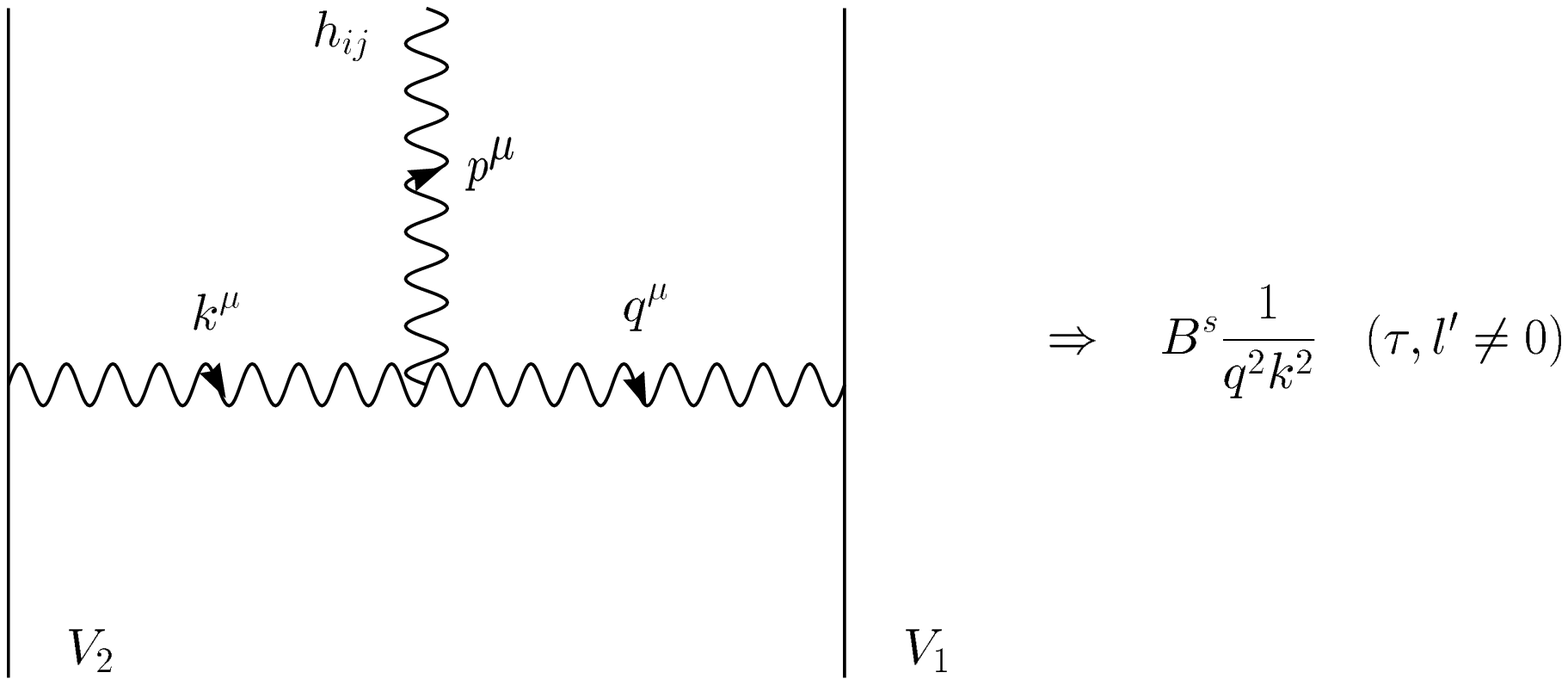,width=350pt}}
\label{fig3}
\end{figure}
\vskip -15pt
\begin{figure}[h]
\centerline{\psfig{figure=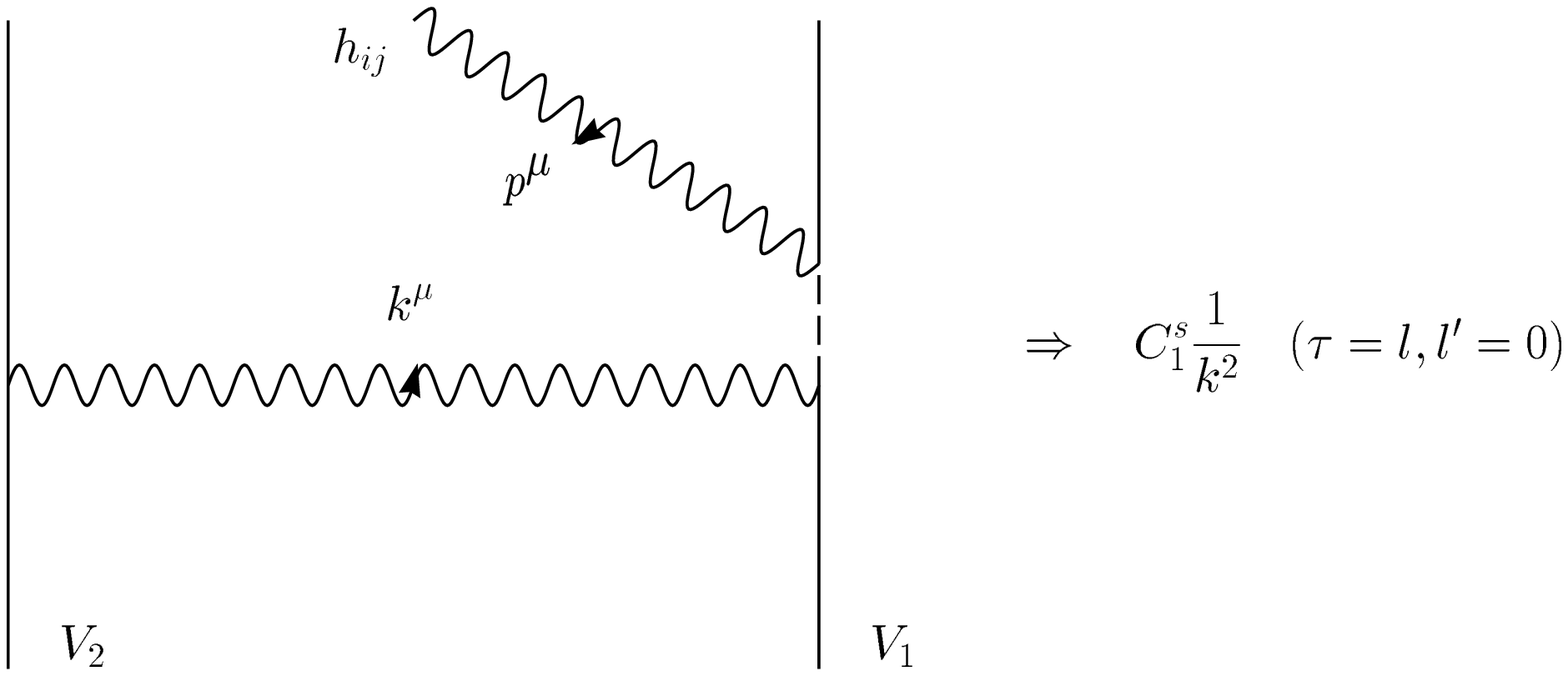,width=362pt}}
\label{fig4}
\end{figure}
\vskip -15pt
\begin{figure}[h]
\centerline{\psfig{figure=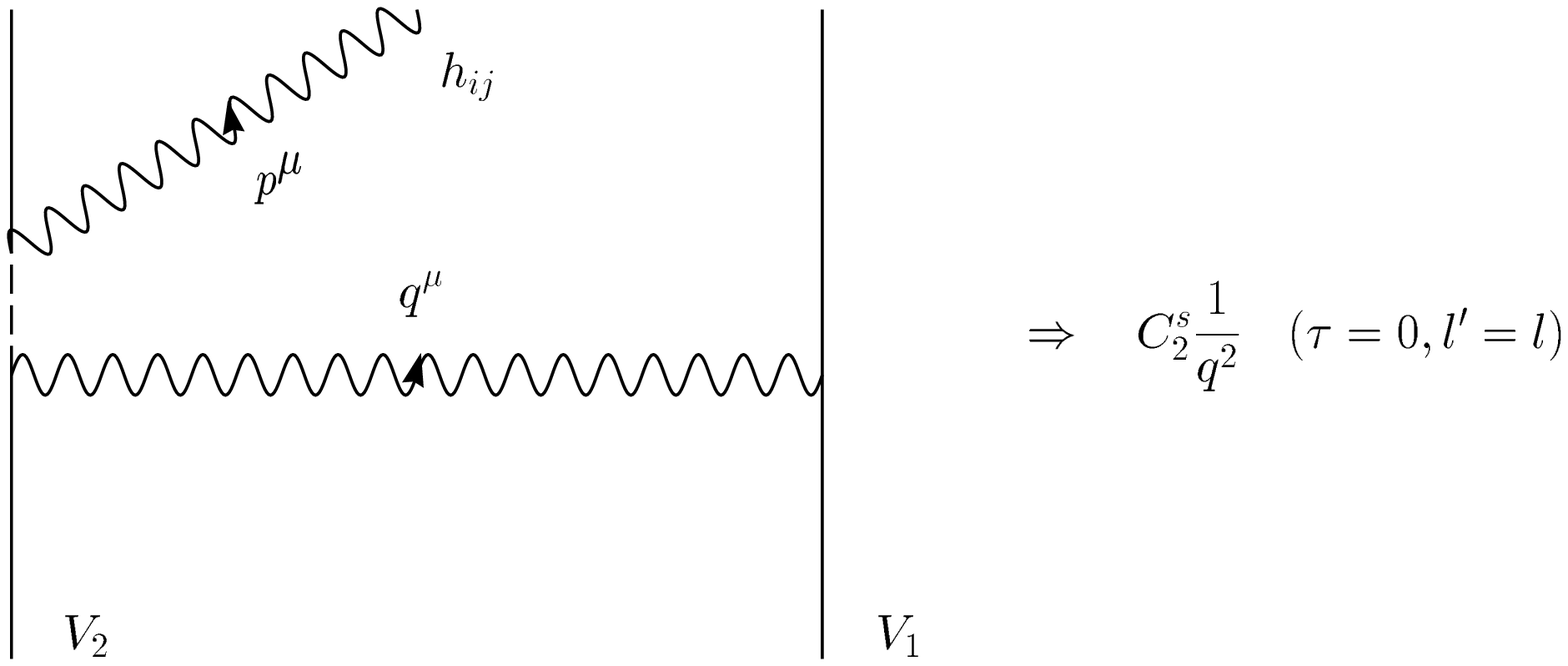,width=362pt}}
\label{fig5}
\end{figure}
\vskip 90pt $\quad$

The $B^s$ factor corresponds to an annihilation process occurring far away 
from both branes, whereas the $C_1^s$ and $C_2^s$ factors correspond to
absorption-emission bremsstrahlung-like processes occurring on the first and
the second brane respectively.

\subsection{Large distances}

It is interesting to compare the string theory results to a field theory
computation in the limit of large impact parameters $\vec b$. 

For the axion and the dilaton, there is no coupling in supergravity allowing 
the emission process, and therefore the vanishing of the string amplitude is 
understood. For the annihilation term of the graviton, there are three 
possible diagrams in supergravity, involving the exchange of vectors,
dilatons and gravitons. Their contributions in the eikonal approximation are
\beqa
\al B^{NS}_{\phi} \sim -a^2  h_{ij} k^i k^j \;, \nn \\
\al B^{R}_{V_\mu} \sim e^2 \left[\cosh v h_{ij} k^i k^j 
- p \sinh v h_{i1} k^i \right] \;, \\
\al B^{NS}_{g_{\mu\nu}} \sim -M^2 \left[\cosh 2v h_{ij} k^i k^j 
- 2p \sinh 2v h_{i1} k^i + 2 p^2 \sinh^2 v h_{11} \right] \;. \nn
\eeqa

The annihilation part of the string amplitude in the various compactification 
schemes is instead the following:

\noindent
{\bf 0-brane: untwisted sector $\&$ 3-brane on $T_2 \otimes T_4/Z_2,T_6$}

\noindent
One finds an exponential enhancement from the partition functions in the 
NSNS sector,
$$
Z^{R+} - Z^{NS+} + Z^{NS-} \rightarrow 16 \cosh v - 4 \cosh 2v - 12 \;,\;\;
Z^{NS+} + Z^{NS-} \rightarrow 2 e^{2\pi l} \;,\nn
$$
and in the final result we recognize a cancellation of the leading order
between the RR vector and the NSNS dilaton and graviton exchange:
\beqa
\al B^{R}_{grav} = 4 \left[\cosh v h_{ij} k^i k^j 
- p \sinh v h_{i1} k^i \right] \;, \nn \\
\al B^{NS}_{grav} = -\left[\cosh 2v h_{ij} k^i k^j 
- 2p \sinh 2v h_{i1} k^i + 2 p^2 \sinh^2 v h_{11} \right] \nn \\
\al \qquad \qquad - 3 h_{ij} k^i k^j \;, \nn \\
\al \Rightarrow B_{grav} \sim V^4 h_{ij} k^i k^j 
+ V^3 p h_{i1} k^i + V^2 p^2 h_{11} \;.
\eeqa

\noindent
{\bf 0-brane: twisted sector}

\noindent
In this case, there is no enhancement in the NSNS sector,
$$
Z^{R+} - Z^{NS+} - Z^{NS-} \rightarrow 4 \cosh v - 4 \;,\;\;
Z^{NS+} - Z^{NS-} \rightarrow 0 \;, \nn
$$
and the cancellation of the leading order occurs between the RR vector and
NSNS dilaton exchange: 
\beqa
\al B^{R}_{grav} = \left[\cosh v h_{ij} k^i k^j 
- p \sinh v h_{i1} k^i \right] \;, \nn \\
\al B^{NS}_{grav} = -h_{ij} k^i k^j \;, \nn \\
\al \Rightarrow B_{grav} \sim V^2 h_{ij} k^i k^j 
+ V p h_{i1} k^i + V^2 p^2 h_{11} \;.
\eeqa

\noindent
{\bf 3-brane on $T_6/Z_3$}

\noindent
In this case there is again an enhancement in the NSNS sector,
$$
Z^{R+} - Z^{NS+} + Z^{NS-} \rightarrow 4 \cosh v - 4 \cosh 2v \;,\;\;
Z^{NS+} + Z^{NS-} \rightarrow 2 e^{2\pi l} \;, \nn
$$
and the cancellation is between the RR vector and the NSNS graviton exchange:
\beqa
\al B^{R}_{grav} = \left[\cosh v h_{ij} k^i k^j 
- p \sinh v h_{i1} k^i \right] \;, \nn \\
\al B^{NS}_{grav} = -\left[\cosh 2v h_{ij} k^i k^j 
- 2p \sinh 2v h_{i1} k^i + 2 p^2 \sinh^2 v h_{11} \right] \;, \nn \\
\al \Rightarrow B_{grav} \sim V^2 h_{ij} k^i k^j 
+ V p h_{i1} k^i + V^2 p^2 h_{11} \;.
\eeqa

The patterns of cancellation in the various cases confirm 
the interpretation in terms of low energy supermultiplets coming from the 
computation of the potential.

\noindent
{\bf Collinear emission}

\noindent
In the case of collinear emission, that is for $\theta = 0$, the results 
simplify a lot and one finds
\beq
B_{grav} \sim V^n h_{ij} k^i k^j \;,\;\; C_{1grav} = C_{2grav} = 0 \;,
\eeq
with $n=2,4$ depending on the amount of supersymmetry left over. Also, in this
case the contributions $C_1$ and $C_2$ from bremsstrahlung-like processes 
vanish identically. 

\subsection{Radiated energy}

To conclude, let us compute the average energy radiated when two D-branes 
pass each other at impact parameter $\vec b$. This is given by
$$
<p> \sim \int \frac {d^3 \vec p}{p} p \left|{\cal A}\right|^2 \;.
$$

For $\theta = 0$ and $V \ll 1$ one obtains
$$
{\cal A} \sim V^{n-1} g_s l_s f(\frac {p \cdot b}{V}) 
e^{- \frac {p \cdot b}{V}} \;,
$$
where $f$ is a slowly varying function and $n = 2, 4$. 
Notice that the emission is exponentially suppressed for 
$p \sim p_{max} = V/b$. By dimensional analysis one finds finally
\beq
<p> \sim g_s^2 l_s^2 \frac {V^{1 + 2n}}{b^3} \;. 
\eeq

\vskip 5pt
\noindent
{\Large \bf Acknowledgments}
\vskip 5pt

Work partially supported by EEC contract ERBFMRX-CT96-0045.

\end{document}